\documentclass[reprint,amsmath,amssymb,pre,superscriptaddress]{revtex4-1}
\usepackage{graphicx}
\usepackage{dcolumn}
\usepackage{bm}
\usepackage[utf8]{inputenc}

\begin{document}

\title{Coarse Graining the Dynamics of Heterogeneous Oscillators \\ in Networks with Spectral Gaps}

\author{Karthikeyan Rajendran}
\email{krajendr@princeton.edu}
\homepage{http://arnold.princeton.edu/~krajendr/}
\affiliation{Department of Chemical and Biological Engineering, Princeton University, Princeton, New Jersey 08544, USA}
\author{Ioannis G. Kevrekidis}
\email{yannis@princeton.edu}
\homepage{http://arnold.princeton.edu/~yannis/}
\affiliation{Department of Chemical and Biological Engineering, Princeton University, Princeton, New Jersey 08544, USA}
\affiliation{Program in Applied and Computational Mathematics (PACM), Princeton University, Princeton, New Jersey 08544, USA}

\date{\today}

\begin{abstract}
We present a computer-assisted approach to coarse-graining the evolutionary dynamics of a system of {\em nonidentical}
oscillators coupled through a (fixed) network structure.
The existence of a spectral gap for the coupling network graph Laplacian suggests that the graph dynamics may quickly become low-dimensional.
Our first choice of coarse variables consists of the components of the oscillator states
--their (complex) phase angles-- along the leading eigenvectors of this Laplacian.
We then use the equation-free framework \cite{Kevr03equation-free}, circumventing the derivation of explicit coarse-grained equations,
to perform computational tasks such as coarse projective integration, coarse fixed point and coarse limit cycle computations.
In a second step, we explore an approach to incorporating oscillator heterogeneity in the coarse-graining process.
The approach is based on the observation of fast-developing correlations between oscillator state and oscillator intrinsic
properties, and establishes a connection with tools developed in the context of uncertainty quantification.

\end{abstract}


\maketitle

\section{\label{sec:intro}Introduction}

The term {\em oscillator} is typically used to denote any physical system which, operating on its own (independent of neighboring oscillators),
exhibits limit cycle behavior.
When such oscillators are coupled to each other, they can spontaneously synchronize with each other.
A simple, yet truly powerful model describing synchronization in oscillator assemblies is the Kuramoto model \cite{Kura84chemical},
which has been successfully used in many biological \cite{Cruz05computer,Winf67biological}, chemical \cite{Ertl91oscillatory},
physical \cite{Wies96synchronization} and social \cite{Neda00sound} contexts.
This model for coupled phase oscillators and its variations have been widely studied in the literature \cite{Aceb05kuramoto}.
Under specific conditions, it has been observed to exhibit complex behavior \cite{Lain09dynamics}.
While extensive work has been performed for all-to-all coupled oscillators,
real-world assemblies of oscillators are seldom globally connected to one other.
Spiking neurons, for instance, are connected by a complex network structure; synchronization of such
neuronal systems has been modeled using the Kuramoto model
modified to account for the network topology \cite{Cumi07generalising}.
Kuramoto oscillators with structured underlying network topologies are increasingly being investigated
in the literature (e.g. \cite{Are08synchronization,Li08synchronization,Mori09synchronization}).

We consider a generic system of {\em non-identical} phase oscillators connected by a network structure, and explore
the computer-assisted reduction of the system dynamics.
Coarse-graining is feasible when there is an inherent {\em separation of time scales} in the system, i.e.,
when constituent processes of the system dynamics occur at very different rates.
Networks with spectral gaps (big jumps in the eigenspectrum of their graph Laplacian)
can endow the coupled oscillator dynamics with this kind of time scale separation.
Our illustrative example is a simple network structure containing a spectral gap after a
relatively small number of eigenvalues (sorted in ascending order) of the graph Laplacian.
The small number of leading eigenvalues before the spectral gap endows the system with {\em low-dimensional} long-term dynamics.
The eigenvectors associated with these eigenvalues (corresponding to ``slow modes") are used to define the coarse variables useful
in model reduction.
Such coarse variables take into account the network structure, but do not account for the fact that the oscillators
in the network are {\em non-identical} in terms of their angular frequencies.
We will discuss how this additional heterogeneity (intrinsic to the oscillators, as opposed to the heterogeneity associated with their
coupling connections in the network) can also be accounted for in the selection of a set of coarse variables.
Once appropriate coarse observables are identified, one typically obtains a reduced set of equations (approximately) describing the evolution of these observables.
In this paper we will circumvent this step using the so-called {\em equation-free} framework \cite{Kevr03equation-free}; in this
approach, short bursts of detailed system simulation are used to estimate the coarse time-derivatives (actions of coarse Jacobians etc.)
required to compute solutions with the coarse variables.
The use of this approach is illustrated in more detail in the Appendix.
The remainder of this paper is structured as follows:
Section~\ref{sec:dyn} describes our illustrative example and outlines its relevant dynamic behavior.
Section~\ref{sec:cg} discusses possible approaches to coarse-graining the system dynamics,
focusing on the selection of appropriate coarse variables (observables).
A first round of results of our coarse-grained computations is presented in Section~\ref{sec:res};
a quick review of the the equation-free framework employed for these computations is relegated to the Appendix.
Sections~\ref{sec:het}~and~\ref{sec:hetc} focus on the heterogeneity of the intrinsic oscillator frequencies,
its effect on their states, and present an approach to account for these effects in coarse-graining.
Section~\ref{sec:conc} concludes with a summary and discussion of possible generalizations of the methods.

\section{\label{sec:dyn}System dynamics}

Our illustrative example is a network of oscillators with a single state variable ({\em phase}) associated with each oscillator.
These phases evolve based on the Kuramoto equations, taking into account the particular connectivity structure:

\begin{equation}
\frac{d\theta_i}{dt}=\omega_i+\frac{K}{N}\sum_{j=1}^NA_{ij}sin(\theta_j-\theta_i), ~1\leq i \leq N.
\label{eq:kura}
\end{equation}

\begin{figure}[b]
\includegraphics[width=0.4\textwidth]{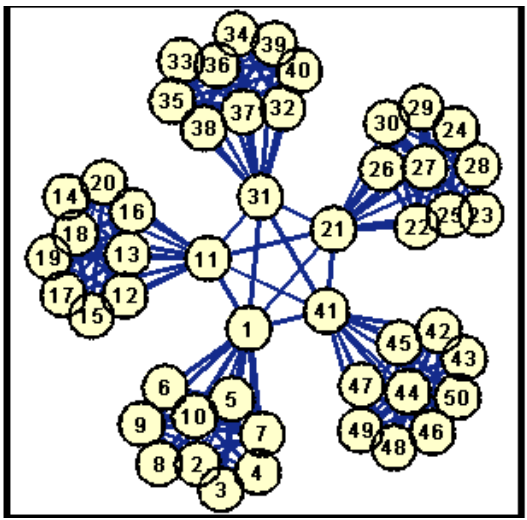}
\caption{\label{fig:graph}A sample graph with a spectral gap, $\mathcal{G}^{(50,5)}$, created using the procedure described in the text.
(The image was created using the graphlayout package for MATLAB written by Matthew Dunham, University of British Columbia.)
}
\end{figure}

Here, $N$ is the total number of oscillators in the system,
$\theta_{i}$ and $\omega_{i}$ are the phases and the intrinsic angular frequencies of the individual oscillators,
and $K$ is the coupling strength, measuring the influence of every oscillator on the oscillators connected to it.
The matrix $A$, the adjacency matrix defining the network structure connecting the oscillators,
is defined as follows: $A_{ij}  = 1$ if oscillators $i$ and $j$ communicate with each other and $A_{ij}  = 0$ otherwise.

\begin{figure}[b]
\includegraphics[width=0.5\textwidth]{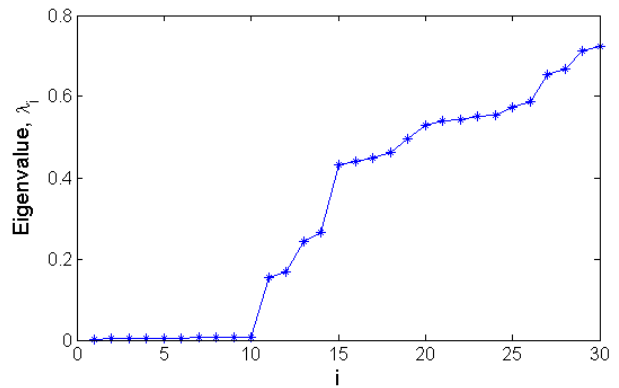}
\caption{\label{fig:leig}The first 100 eigenvalues of the graph Laplacian corresponding to $\mathcal{G}^{(500,10)}$.}
\end{figure}

\begin{figure*}
\includegraphics[width=1\textwidth]{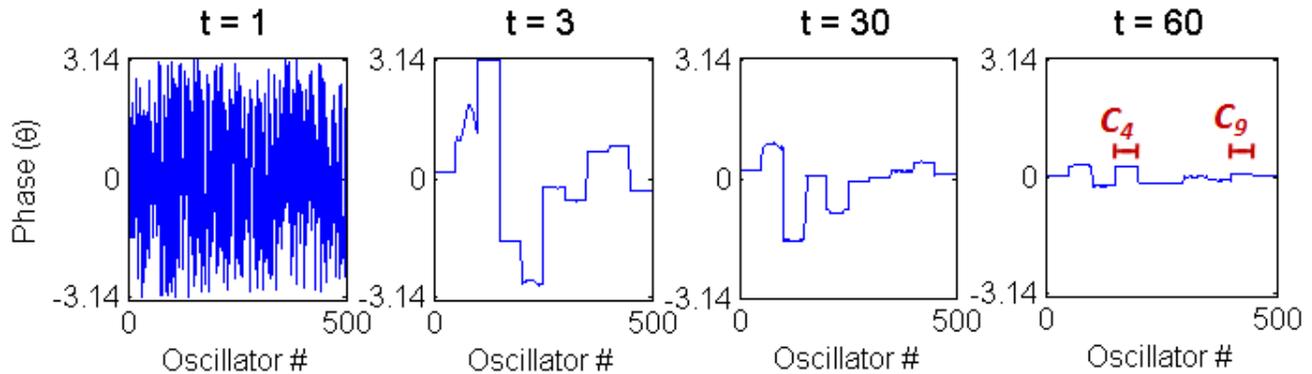}
\caption{\label{fig:phaseev}The temporal evolution of the phases of the oscillators at a coupling strength of $K=0.5$.
The oscillators in a couple of representative communities (the fourth, $C_4$, and ninth, $C_9$)
are marked in the last plot of the sequence.
}
\end{figure*}

\begin{figure}[b]
\includegraphics[width=0.5\textwidth]{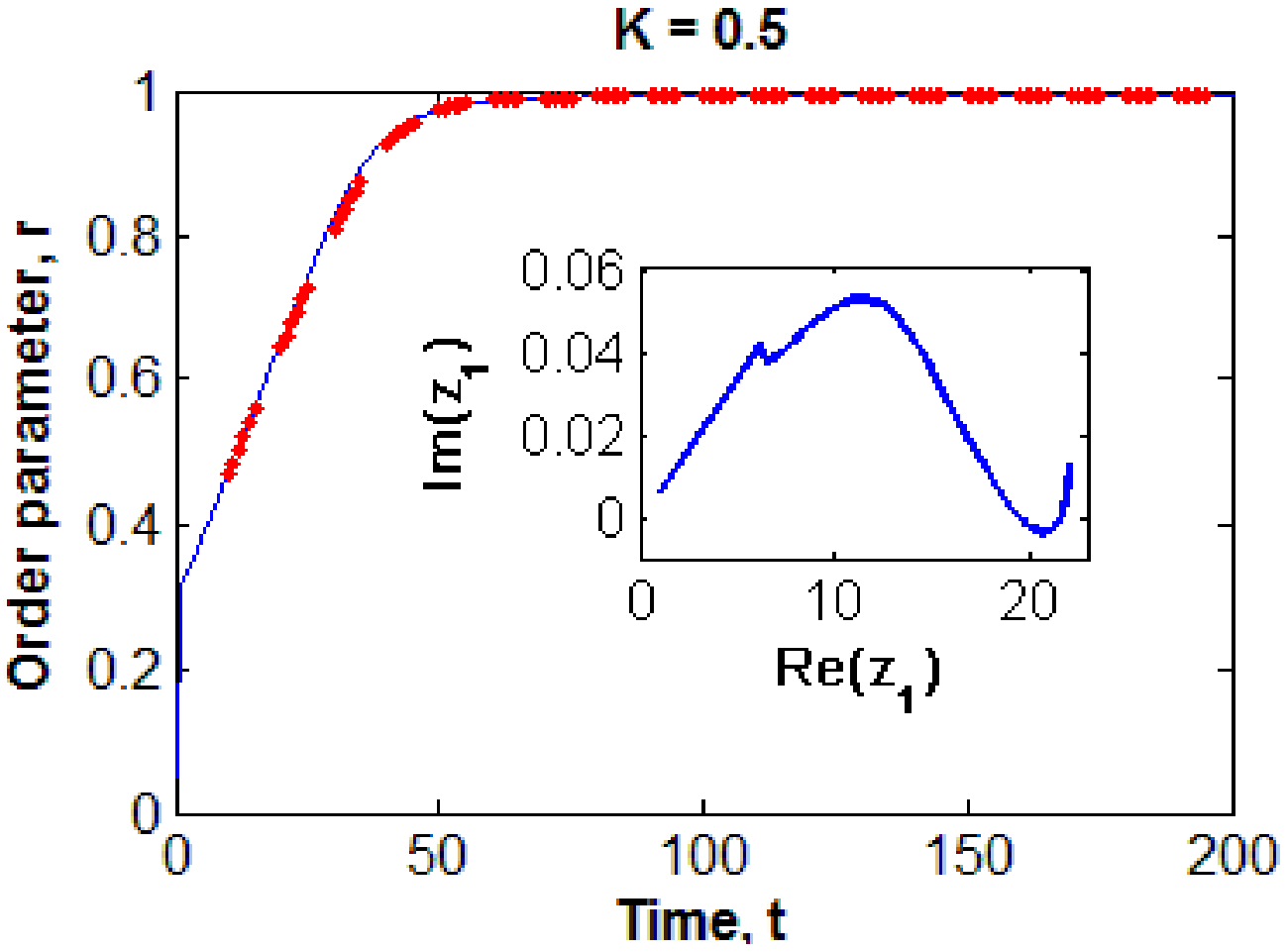}
\caption{\label{fig:order1} The evolution of the Kuramoto order parameter over time at a coupling strength of $K=0.5$.
The results from direct simulation are the solid lines (blue) while the results from coarse projective integration
($5$ time steps for simulation and $5$ for projection) are the dots (red).
(The thickness of the plotted dots make the projection step appear shorter).
The phase portrait in terms of the real and imaginary components of the first coarse variable (see Eq.~\ref{eq:coarse2}) is shown in the inset.}
\end{figure}

\begin{figure}[b]
\includegraphics[width=0.5\textwidth]{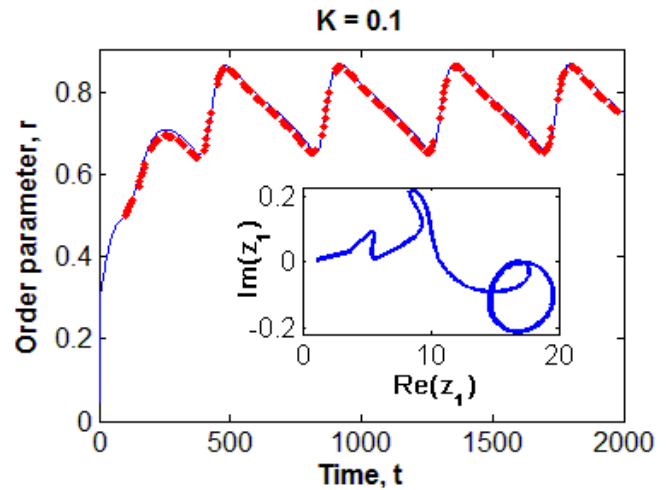}
\caption{\label{fig:order2} The evolution of the Kuramoto order parameter over time at a coupling strength of $K=0.1$.
The results from direct simulation are the solid lines (blue) while the results from coarse projective integration
($25$ time steps for simulation and $25$ for projection) are the dots (red).
(The thickness of the plotted dots make the projection step appear shorter).
The phase portrait in terms of the real and imaginary components of the first coarse variable (see Eq.~\ref{eq:coarse2}) is shown in the inset.}
\end{figure}

For our simulations, we use $500$ oscillators with their intrinsic frequencies sampled from a Gaussian distribution with mean $0$ and standard deviation $1/15$;
we will discuss the possibility of different distributions in Section~\ref{sec:conc}.
For the underlying connectivity we built a network with a spectral gap, which was observed to lead to low-dimensionality in the long-term system dynamics;
this provides the motivation for coarse-graining.
The target graph for our illustration was created from a collection of $m$ subgraphs (communities) each containing $s$ nodes;
the total number of nodes (oscillators) in the final network was $N=m \times s$.
%
%
Each subgraph was created using the Watts-Strogatz model \cite{Watt98collective}, which contains 2 parameters,
$k$ - the average degree of the nodes and $p$ - the probability of rewiring.
The values of $k$ for the $m$ subgraphs were assigned by uniformly sampling an even number in the interval [14,38]
(corresponding to an average degree of approximately 25-75\% of the total number of nodes in the sub-graph).
The values of $p$ for the individual subgraphs were sampled uniformly in a log scale between 0.001 and 1
(i.e., the values of $log_{10}~p$ are sampled uniformly between -3 and 0).
The Watts-Strogatz model was chosen to create the constituent subgraphs because it creates graphs ranging from
Poisson degree distribution (random) to power law distribution (scale-free) depending on the parameter $p$.
Once all the sub-graphs (or communities) are created, a node is randomly chosen from each of the communities to be its {\em leader}.
Now, all the $m$ leaders are connected to each other resulting in a complete network of leaders (with ${{m}\choose{2}}$ edges).
We thus arrive at a connected graph, $\mathcal{G}^{(N,m)}$ with $m$ communities and $N$ nodes in total;
a sample resulting graph is shown in Fig.~\ref{fig:graph} for the case of $m=5$ and $s=10$.
In our simulations, we use a graph $\mathcal{G}^{(500,10)}$ created using the same procedure with $m=10$ and $s=50$.
The normalized Laplacian of the graph, denoted by $L$, is defined as:
\begin{equation}
L_{ij} := \left\{
\begin{array}{cl}
1 & \text{if } i=j \text{ and } d_{i} \neq 0,\\
-1/\sqrt{d_{i}d_{j}} & \text{if } i \neq j \text{ and } A_{ij}=1,\\
0 & \text{otherwise}
\end{array} \right.
\label{eq:lap}
\end{equation}
where $d_i$ is the degree of node $i$.

The normalized Laplacian (in this paper, the term Laplacian should always be
taken to mean the normalized graph Laplacian) corresponding to our graph $\mathcal{G}^{(500,10)}$
was computed and its first few eigenvalues (arranged in the ascending order) are plotted in Fig.~\ref{fig:leig};
there is a clear gap in the spectrum after the 10th eigenvalue.
We perform direct simulations of the phase oscillator model (Eq.~\ref{eq:kura}) at different values of coupling strength,
with the network topology and oscillator frequencies chosen as described above; the initial phases of the oscillators are
sampled from a uniform distribution between $-\pi$ and $\pi$.
All our results are reported after the instantaneous average system phase $\arg ( \sum_{j} e^{i\theta_j} ) $ has been subtracted
(i.e., in a frame that rotates along with the average system phase).
At sufficiently large values of coupling strength $K$ we observe (as expected) that the oscillators spontaneously
synchronize their frequencies, and their phases ``lock" at steady state.
Representative phase evolution at such a high coupling strength $K=0.5$ is shown at successive time steps in Fig.~\ref{fig:phaseev};
note how the community structure of the oscillators quickly becomes visually apparent in the figure.
A quantitative measure of {\em phase synchronization} (or coherence in an oscillator population), the
so-called {\em order parameter} has been defined as:
\begin{equation}
r=\left\| \frac{1}{N}\sum_{j=1}^N e^{i\theta_j} \right\|.
\label{eq:order}
\end{equation}
Its values can range between $0$ and $1$.
The higher the value of the order parameter, the higher the degree of synchronization - a value of $1$ indicates the state where
all oscillators have the exact same phase.
The evolution of this order parameter is shown at a coupling strength of $K = 0.5$ as solid lines in Fig.~\ref{fig:order1}.
As expected (and confirmed computationally) the steady state value of the order parameter decreases with coupling strength until
a critical value of $K_{c}$; below this critical value the oscillators do not synchronize any more, and
limit cycle oscillations are observed initially (in parameter space), resulting from a supercritical Hopf bifurcation
at $K_{c}$.
The evolution of the order parameter at $K=0.1$ depicted by the solid lines in Fig.~\ref{fig:order2} exhibits such steady limit cycle oscillations.

\section{\label{sec:cg}Coarse-graining}

Our objective is to develop and implement a computer-assisted coarse-grained model of our illustrative coupled oscillator system.
The most important step in coarse-graining probably consists of the selection of appropriate coarse observables: a reduced set
of variables in terms of which a useful {\em closed} description can be obtained.
Given suitable coarse variables, it is sometimes possible to derive the reduced equations analytically (in closed form).
If, however, the closures required to ``write down" the reduced equations in closed form are not known, or cannot be
easily approximated, the so-called equation-free framework \cite{Kevr03equation-free,Kevr04equation-free:} can be used to
{\em computationally} implement the reduced model, circumventing its explicit derivation.
A brief description of the main points of this modeling/computation framework for complex systems can be found in the Appendix.

In order to motivate our selection of coarse variables, we first identify collective features in the detailed simulation results.
Consider the temporal evolution of the oscillator phases as shown in Fig.~\ref{fig:phaseev} (for $K=0.5$).
In our network $\mathcal{G}^{(N,m)}$, the first $s (=N/m)$ oscillators belong to the first community, the next $s$ to the second community and so on;
we define $C_k$ as the set containing the indices of all the oscillators in the $k^{th}$ community:
\begin{equation}
C_{k\in[1,m]} = \{(k-1)s+1,(k-1)s+2,...ks\}.
\label{eq:comm}
\end{equation}

Two oscillators within the same community are connected ``more tightly" (they share more common neighbors) than oscillators in different communities.
We {\em observe} in the simulations that the phases of all the oscillators within a community
synchronize with each other at much shorter time scales than those over which the entire network synchronizes.
This suggests that, because of the {\em construction} of our network topology, its structure can help rationalize the
selection of coarse variables appropriate for the {\em long-term} dynamics, after initial transients quickly
die out.
That this separation of time scales leads to low-dimensionality in the system state can be seen in Fig.~\ref{fig:phaseev}:
the randomly initialized individual oscillator phases (our ``microscopic" or ``fine scale" variables, $U$ in
equation-free notation)
quickly evolve to ``respect" the coarse community structure of the network.
%
%

As a result, the following possibilities for coarse variables suggest themselves:

\subsection*{Option 1: Average phase in each community}

An obvious choice for a set of coarse variables, $u^{(1)},$ for our example is to use a single common (time-varying) phase angle for each community.
The {\em restriction operation} (fine states to coarse states in equation-free language)
is then defined by assigning the average phase, $\overline{\theta}_k$, of all the oscillators in the $k^{th}$ community
as the single common phase of that community.
The {\em lifting operation} (coarse states to consistent fine ones) is implemented by assigning this common phase
as the phase angle of all the oscillators in that community.
\begin{eqnarray}
& \overline{\theta}_k = \frac{1}{s} \sum_{j\in C_k}^{} \theta_j,
\label{eq:avg}
\\
& u^{(1)} = \{\overline{\theta}_{k \in [1,m] }\}.
\label{eq:coarse1}
\end{eqnarray}

This apparently intuitive coarse variable selection suffers from two drawbacks.
Firstly, partitioning the oscillators into different communities (``clustering" \cite{Newm06finding,Nadl06diffusion,Gfel08spectral}) is nontrivial for a general
network structure (even though -due to the particular construction- it was easy for our example).
Even when community structures can be identified, however, this set of coarse variables does not take
into account the differences {\em between} the different communities and the structure {\em within} the communities.
This suggests an alternative set of coarse variables that uses the graph Laplacian of the network.

\begin{figure}[t]
\includegraphics[width=0.5\textwidth]{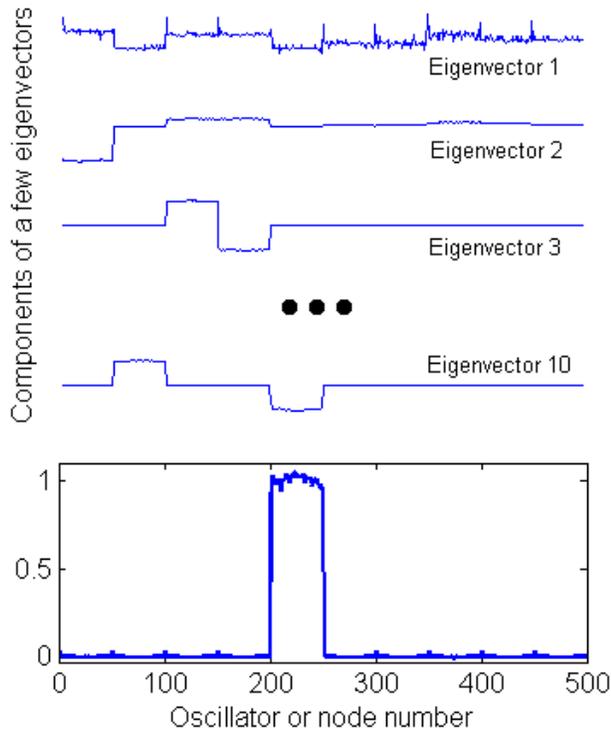}
\caption{\label{fig:evect}The components of a few eigenvectors of the Laplacian of the graph $\mathcal{G}^{(500,10)}$ are plotted in arbitrary units.
The plot (blue line) at the bottom corresponds to a vector that is a linear combination of the first 10 eigenvectors;
it is clearly localized to oscillators in the fifth community only.
The ten linear combination coefficients are [2.066, 0.259, 0.004, 0.273, -0.323, 0.222, -1.151, 0.017, 4.385, -4.982].}
\end{figure}

\subsection*{Option 2: Projection to a (truncated) Laplacian eigenbasis}

Consider the normalized Laplacian matrix, defined in Eq.~\ref{eq:lap}, for the graph $\mathcal{G}^{(500,10)}$.
Let $\lambda_j$ be the $j^{th}$ eigenvalue and $\textbf{v}_j$ the corresponding normalized eigenvector of the graph Laplacian.
From Fig.~\ref{fig:leig}, it can be seen that the first 10 eigenvalues are well separated from the rest (in other words, a spectral gap exists).
The components of the eigenvectors of the graph Laplacian corresponding to the ten smallest eigenvalues
are plotted as connected dotted lines in Fig.~\ref{fig:evect}.
These eigenvectors embody, in an alternative way, the coarse community structure of the entire network.
Linear combinations of these eigenvectors can be used to approximately represent any one of the individual communities
(as an example, a linear combination of the first 10 eigenvectors whose support lies (approximately) only
on the oscillators in the fifth community is shown as a thick solid line in Fig.~\ref{fig:evect}).
When only a few eigenvectors capture the presence and structure of the different communities,
they form a suitable basis to represent the long term dynamics of the system.
A comparison of Figs.~\ref{fig:phaseev} and ~\ref{fig:evect} also suggests that a linear
combination of these eigenvectors is likely to represent well the long term evolving states of the phase oscillators.
In other words, after a short initial transient, the system becomes attracted to a {\em low-dimensional manifold}
on which the individual oscillator phases can be represented using
a lower dimensional basis formed by the first few (here 10) Laplacian eigenvectors of the network;
these eigevectors {\em parametrize} the low-dimensional manifold.

The use of the graph Laplacian in constructing (coarse) observables is, of course, not new
\cite{Aren06synchronization,Aren06synchronizationa,McGr07analysis,McGr08laplacian}.
Below we will {\em use} these variables not only as observables, but as the means to implement
accelerated coarse-grained computations with the model.
There is a clear analogy between such observables and the use of the finite Fourier transform in
solving initial-boundary value problems: instead of eigenfunctions of diffusion in {\em physical space}
we have eigenvectors of the Laplacian {\em on a graph}; instead of evolution equations for time-dependent
Fourier mode amplitudes we envision evolution equations for time-dependent components of the system phases
on the first few graph Laplacian eigenvectors.
Finally, it is also worth pointing out that these eigenvectors are also the eigenvectors of the
linearization of the model around the uniform solution of a ``nearby" problem: that of coupled {\em identical}
oscillators \cite{Aren06synchronization,Aren06synchronizationa}.
We start by defining an $N \times 1$  vector of {\em complex} phase angles of the oscillators, \mbox{\boldmath$\Theta$}:
\begin{equation}
\Theta_j = e^{i\theta_j};~j \in [1,N].
\label{eq:complex2}
\end{equation}

The phase angles should be defined modulo $2 \pi$; this complex phase vector correctly represents the phase variable on a unit circle
(described by $sin~\theta$ and $cos~\theta$).
We now choose as coarse variables ($u^{(2)}$) the components, $z_j$, of this phase vector,
\mbox{\boldmath$\Theta$}, along the direction of the first ten eigenvectors of the graph Laplacian.
\begin{eqnarray}
& L\textbf{v}_j = \lambda_j \textbf{v}_j; ~j \in [1,N],
\label{eq:evect}
\\
& u^{(2)} = \{z_{j \in [1,m]} = \textbf{v}_j^{T} \mbox{\boldmath$\Theta$} \}.
\label{eq:coarse2}
\end{eqnarray}
This projection is our {\em restriction}, while
translation between the fine description (phases of all the $N$ oscillators)
and the coarse description is governed by Eq.~\ref{eq:lift2}, our {\em lifting} operator.
\begin{equation}
\sum_{j=1}^{m} z_j \textbf{v}_j \rightarrow \mbox{\boldmath$\Theta$}.
\label{eq:lift2}
\end{equation}
%
%

A third option for coarse graining, which includes additional heterogeneity considerations is discussed in Sec.~\ref{sec:het}.

\section{\label{sec:res}Computational Results}

\subsection{Coarse projective integration}
Using the set of coarse variables $u^{(2)}$, we accelerated the network simulation using
{\em coarse projective integration} as detailed in the Appendix.
Representative results are shown in Figs.~\ref{fig:order1} and \ref{fig:order2}, reported in the
form of time-series of the order parameter for two values of the coupling strength,
above and below $K_c$ respectively.
For both these cases, the magnitude of the projective step is equal to the duration of the
full simulation used to estimate the coarse time derivatives; that is,
the full system, Eq.~\ref{eq:kura}, is simulated for only $50\%$ of the overall evolution time.
The coarse evolution in both cases clearly follows (in the ``eye norm") the resolved full direct simulation results;
this demonstrates the accuracy of the equation-free approach and indirectly validates our
selection of coarse variables.

\subsection{Coarse fixed point computation}

We performed coarse-fixed point calculations (as outlined in the Appendix)
using both choices of coarse variables and compared it to steady states calculated with the full model.
Given an initial guess of the $10$ coarse variable steady values,
the coarse time-stepper was constructed by lifting, followed by full model simulation
(a representative time-stepper horizon was $\tau = 10$) and restriction.
Fixed points of the coarse time-stepper were arrived at through a Newton-Krylov GMRES iteration
\cite{Saad86gmres:,Kell95iterative}.
To quantify accuracy, we calculated the pairwise linear correlation coefficient between the
detailed (``fine scale") phases of the actual fixed point,
and those lifted from the converged coarse fixed point values for each choice of coarse variables.
The results are reported in Table~\ref{tab:fixed} for three different coupling strengths.
The first(respectively, second) row corresponds to results obtained using $u^{(1)}$(respectively, $u^{(2)}$) as coarse variables.
Clearly, $u^{(2)}$ gives a more accurate coarse description of the system fixed points compared to just using the average phases in the communities ($u^{(1)}$).

\begin{table}
\caption{\label{tab:fixed}Correlation between the detailed phases of the actual fixed point,
and those lifted from the converged coarse fixed point values for each choice of coarse variables.}
\begin{ruledtabular}
\begin{tabular}{c|ccc}
$\rho$ & $K=1$ & $K=0.5$ & $K=0.2$ \\ \hline
Using $u^{(1)}$ & 0.9974 & 0.9975 & 0.9976\\
Using $u^{(2)}$ & 0.9983 & 0.9983 & 0.9983\\
Using $u^{(3)}$ & 0.9994 & 0.9994 & 0.9995\\
\end{tabular}
\end{ruledtabular}
\end{table}

\begin{figure}[b]
\includegraphics[width=0.5\textwidth]{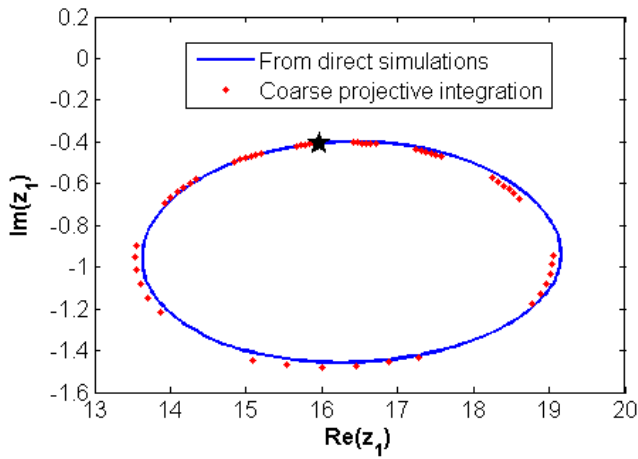}
\caption{\label{fig:limit} A coarse limit cycle computed from direct simulations at a coupling strength of $K=0.1$ is plotted using a solid (blue) line.
The star (black) corresponds to the solution (the point on a Poincar\'e map) obtained using a coarse limit cycle solver.
This point is then used as the initial condition for coarse projective integration (red dots); the coarse trajectory returns to that
point after one period.}
\end{figure}

\subsection{Coarse limit cycle computation}
We have already discussed the existence of stable limit cycle oscillations
below (and close to) the critical coupling strength, $K_c$.
The limit cycle found out from direct simulations for $K=0.1$
is plotted as a solid line in the phase space projection on the real and imaginary parts of $z_1$ in Fig.~\ref{fig:limit}.
We also find a (coarse) point on this limit cycle by locating the (coarse) fixed point of an appropriate (coarse) Poincar\'e map;
the point is represented as a star in Fig.~\ref{fig:limit}.
This point (as well as the period of the coarse limit cycle) is found by solving
(again through Newton-Krylov GMRES)
Eq.~\ref{eq:limit} for the appropriate set of values of the coarse variables $u^{(2)}$;
the Poincar\'e map was defined in terms of the $Re(z_1)$ coarse variable.
In these computations, the full system was simulated for the entire Poincar\'e return time;
but the map, and the Newton fixed point computation were performed in the {\em reduced} space
of the coarse variables.
In an extra validation step, the trajectory around the limit cycle was followed through coarse projective
integration (see Fig.~\ref{fig:limit}), and seen to coincide visually with the (phase space projection of the) full
simulation limit cycle.

These representative computations confirm that computational coarse-graining (with the appropriate selection
of coarse variables) can be used to effectively perform computations with the (explicitly unavailable) coarse-grained
model.
Coarse projective integration, coarse fixed point and limit cycle computations (and also, easily, coarse stability
and continuation computations) can be implemented in the form of computational ``wrappers" around the full
simulation.
The choice of coarse variables (and the associated lifting and restriction steps) form a critical part
of the approach; an improvement on this process is presented below.

\section{\label{sec:het} The effect of oscillator heterogeneity}

\begin{figure*}[!]
\includegraphics[width=0.9\textwidth]{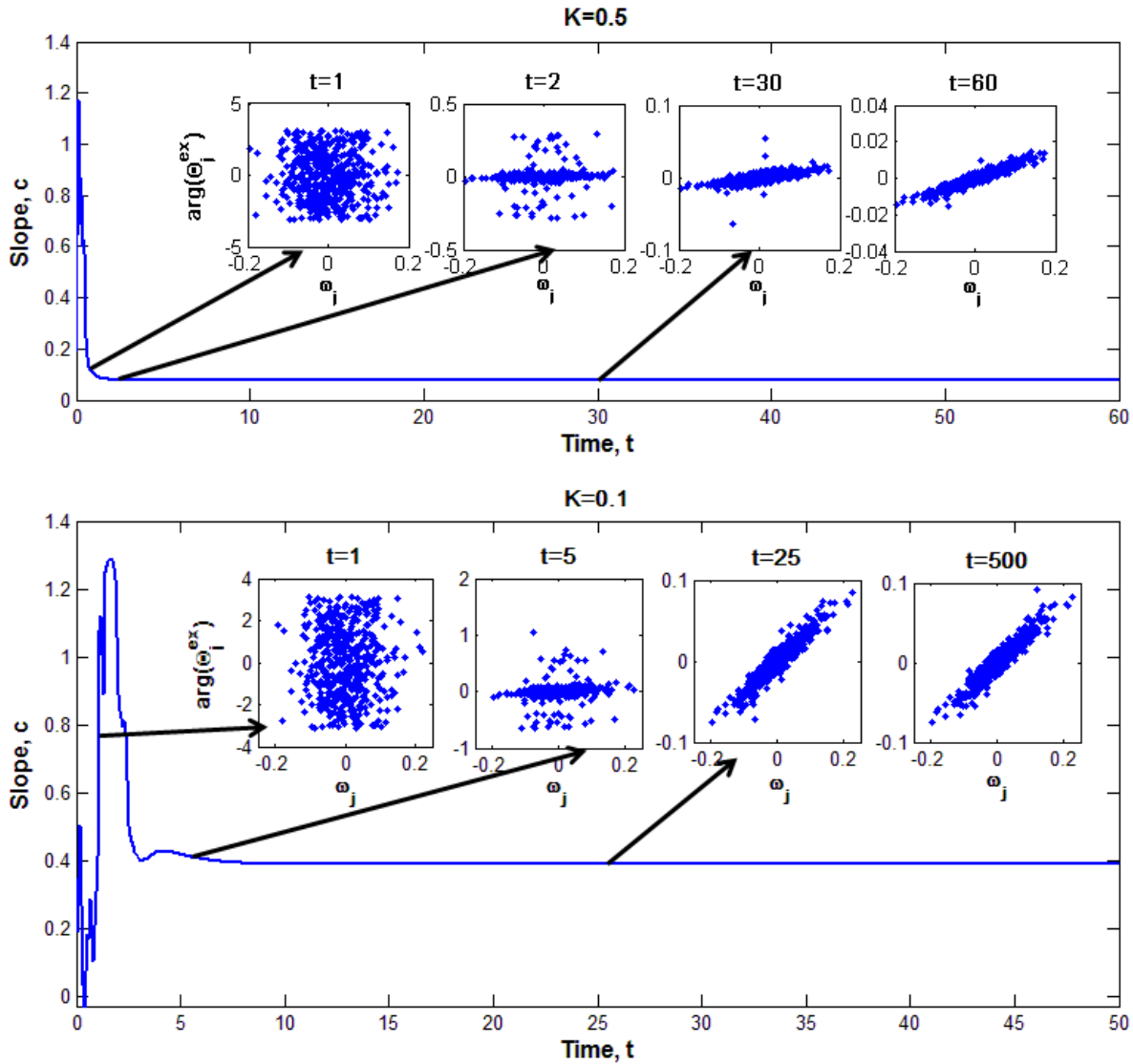}
\caption{\label{fig:slopeT} Evolution of the slope $c$ of the linear fit between excess phase and angular frequency for $K = 0.5$ and $K = 0.1$.
{\em Insets:} Plots of excess phases $arg(\Theta_{j}^{ex}) $ versus oscillator angular frequencies $\omega_j$ at a few representative temporal instances.}
\end{figure*}

\begin{figure*}[!]
\includegraphics[width=1.0\textwidth]{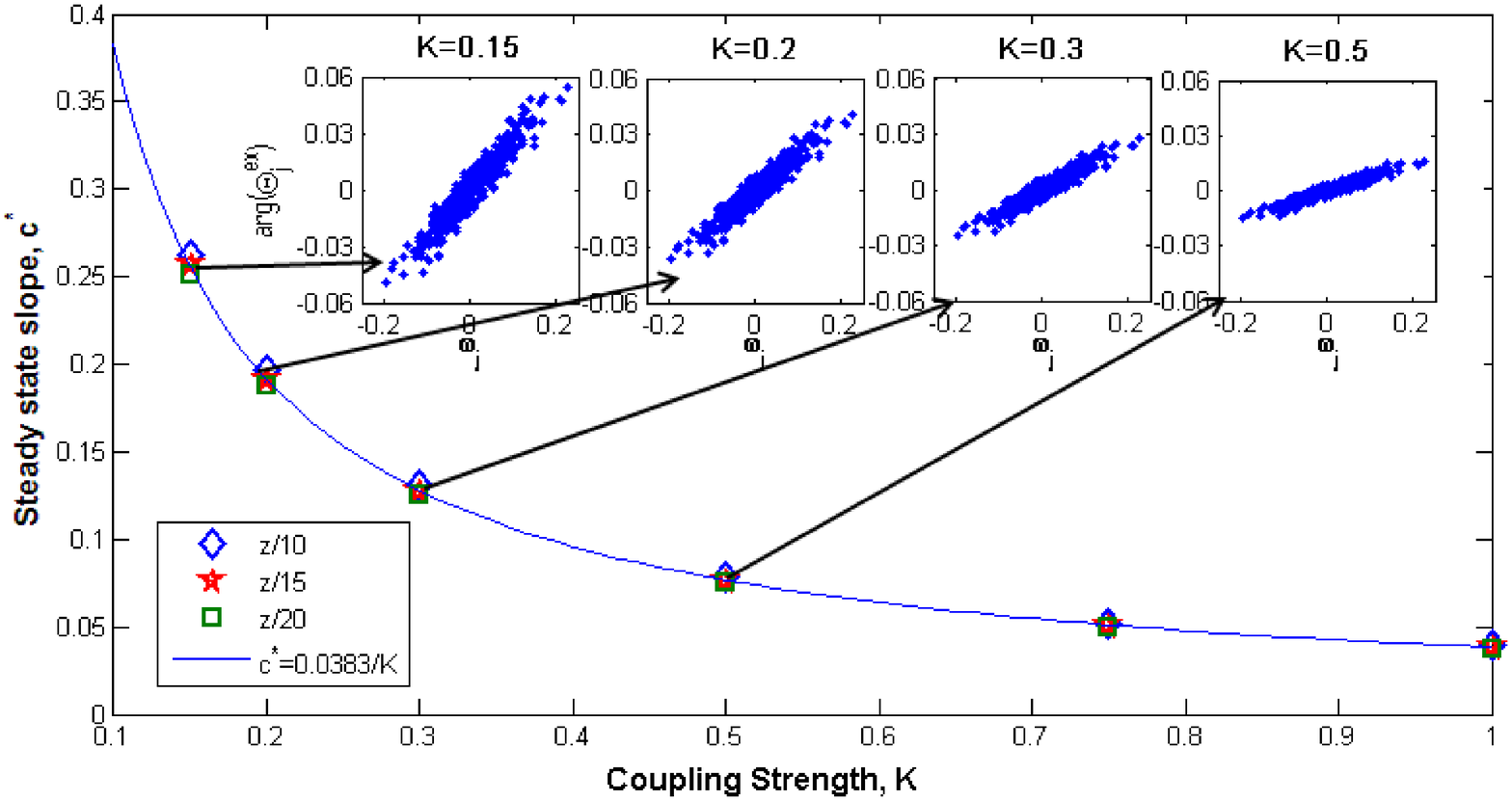}
\caption{\label{fig:slopeK} Steady state values of slope $c^{*}$ of the linear correlation between excess phase
and intrinsic oscillator angular frequency plotted as a function of the coupling strength $K$.
The labels $z/10$, $z/15$ and $z/20$ in the legend denote the three different distributions from which
the intrinsic oscillator frequencies are sampled, $z$ representing a standard normal random variable).
The inverse proportionality can be seen from the curve fit, $c^{*}=0.0383/K$.
{\em Insets:} Plots of excess phases $arg(\Theta_{j}^{ex}) $ versus oscillator angular frequencies $\omega_j$
at steady state for a few values of coupling strength.
}
\end{figure*}

In our coarse graining efforts, so far, we have focused on the structure of the network and ignored the effect of the
heterogeneity in the angular frequencies of the oscillators.
In this section we present a systematic approach to taking this
heterogeneity into account in the coarse-graining procedure.
The motivation comes from the observation, in the literature \cite{moon06coarse}, that for all-to-all coupling
(that is, in the absence of fine network structure) the oscillator phases will, under certain conditions,
become {\em quickly} correlated with their intrinsic frequencies.
We have observed the same phenomenon in our, more structured, networks:
the sequence of insets in Fig.~\ref{fig:slopeT} shows the transient evolution of the (``excess")
phase of the oscillators in our network plotted against their individual frequencies.
The term ``excess phase", $ \mbox{\boldmath$\Theta$}^{ex} $, here refers to the portion of the phase vector that is not captured by the
projection on the first 10 Laplacian eigenvectors; using Eq.~\ref{eq:excess} we plot the complex argument of each oscillator, {\em viz.},
$arg(\Theta_{j}^{ex}) $ against $\omega_j$.
\begin{equation}
 \mbox{\boldmath$\Theta$}^{ex} = \mbox{\boldmath$\Theta$} - \sum_{j=1}^{m} \textbf{v}_j^{T} \mbox{\boldmath$\Theta$} \textbf{v}_j,
\label{eq:excess}
\end{equation}

What we see is that, even when the oscillators are initialized with random phases (in the form of the ``cloud" seen in the first inset),
they very quickly (visibly by time $t=2$ and almost quantitatively by $t=30$) develop a {\em strong, stationary} correlation with the
intrinsic frequencies.
These plots confirm the existence of a strong linear correlation between the ``excess phase" and the angular frequencies of the oscillators.
The evolution of the slope of the best linear fit is plotted in Fig.~\ref{fig:slopeT} (a) and (b) for different values of the coupling strength $K$.

A number of additional observations can be made from these plots.
The time taken for the correlation slope $c$ to approach steady state is {\em much less} than the time scale in which the
order parameters reach steady state (compare with Figs.~\ref{fig:order1} and~\ref{fig:order2}).
Even for the case of $K = 0.1$, corresponding to stable limit cycle oscillations,
the values of $c$ do not vary much once the stable limit cycle is approached.
Fig.~\ref{fig:slopeK} shows how this (quickly achieved) steady state correlation slope varies with the oscillator coupling strength.
For a range of coupling strengths, this steady state value of $c$ is obtained for three different frequency distributions (the oscillator frequencies were sampled from the normal distribution with mean $0$ and standard deviation of $1/10$, $1/15$ and $1/20$
respectively.)
The steady state correlation between excess phase and frequency appears to be
{\em independent} of the range of (or the variance in) the oscillator frequencies.
Fig.~\ref{fig:slopeK} and its insets quantify the dependence of
the correlation on $K$;
%
%
the steady state values of $c$ are seen to decrease with coupling strength as one might expect
(since, at higher coupling strengths, the oscillator phases should exhibit less variance).
In particular, an approximate inverse proportionality is computationally observed
between the coupling strength $K$,
and the steady slope $c^*$, quantified by the following fitted curve:
\begin{equation}
c^{*}=0.0383/K.
\label{eq:slopefit}
\end{equation}

We find it remarkable that our ``decoupled" procedure, which first considers heterogeneity in the network {\em structure},
and only then considers heterogeneity in the {\em oscillator intrinsic properties} gives us so robust features for the network dynamics.
Note  that the constant $0.0383$ applies only for the particular network used in the simulation shown. %
For different network structure realizations, the constant will be different.

Based on these results, we can now integrate the effects of both network structure and oscillator frequency
 distribution in the
coarse-graining of the oscillator phases (in particular, in our {\em lifting operator}).

\section{\label{sec:hetc} Coarse-graining the heterogeneous oscillators}

The discussions in Section.~\ref{sec:het} suggest that capturing the
correlation between excess phase and intrinsic oscillator frequency
can lead to a better set of coarse variables and a more accurate
lifting operator for the coarse-graining process.
For our illustrative example, a single additional variable, the slope
$c$ is, as we will show, sufficient in capturing this correlation.
Before we demonstrate this, however, we note the more general
question: for arbitrary heterogeneity distributions (not just Gaussian
as the one studied here), what is the nature and number of additional coarse  variables necessary to quantitatively account for the frequency heterogeneity?
We will return to this question in the last Section.
Using the single scalar slope $c$ as an additional coarse variable, we define the ``corrected" vector of complex phase angles,
\mbox{\boldmath$\widetilde{\Theta}$}, similar to \mbox{\boldmath$\Theta$}, but now taking the $c\omega_j$ into account:
\begin{eqnarray}
& \mbox{$\widetilde{\Theta}_j$} = e^{i(\theta_j - c \omega_j)};~j \in [1,N],
\label{eq:complex3}
\\
& \sum_{j=1}^{m} z_j \textbf{v}_j \rightarrow \mbox{\boldmath$\widetilde{\Theta}$}.
\label{eq:lift3}
\end{eqnarray}
Our corrected lifting operation, going beyond the first $m$ (here, 10) eigenvectors of the graph
Laplacian, is given by Eqs.~\ref{eq:complex3} and ~\ref{eq:lift3}.
For the corrected restriction operation, the vector of phase angles is initially projected on the
first $10$ graph Laplacian eigenvectors to obtain the $z_j$; the ``excess phases" are then used to estimate the slope $c$ through linear regression.
Our augmented set of coarse variables now reads:
\begin{equation}
u^{(3)} = \{z_{j \in [1,m]} = \textbf{v}_j^{T} \mbox{\boldmath$\widetilde{\Theta}$} ,c\}.
\label{eq:coarse3}
\end{equation}

The results of computational coarse graining with the coarse
variable set $u^{(3)}$ are, as one might expect, qualitatively similar to,
{\em but more accurate than}, the results with the coarse set $u^{(2)}$ presented in Sec.~\ref{sec:res}.

Note that since (as we computationally observed) $c$ quickly approaches
an approximately constant value $c^*$ for each specific system realization, we can
-in studying long term dynamics- {\em fix} its value at $c^*$ and not even consider it as an extra dynamic coarse variable.
The constant value $c^*$ for different coupling strengths can also be inferred from formulas like Eq.~\ref{eq:slopefit}.

Each successive coarse variable choice $u^{(1)}$, $u^{(2)}$ and $u^{(3)}$ clearly includes more information
about the system than the previous one: the coarse variable set $u^{(1)}$ just accounts for the presence of
different communities, $u^{(2)}$ accommodates the structure of the different communities while set $u^{(3)}$
considers the influence of the heterogeneous frequencies of the oscillators as well.
In order to quantify whether this additional information is also meaningful, we compared the results of coarse fixed point analysis
using the three choices of coarse variables.
The results in  Table~\ref{tab:fixed} demonstrate that the additional information included in successive choices of coarse variable
sets actually leads to more accurate computation of the system features (in particular, its coarse steady states).

\section{\label{sec:conc}Conclusions}

We have demonstrated an approach to coarse-graining the computations
of the (long-term) dynamics of networks of coupled heterogeneous oscillators; our approach was based on the equation-free framework,
and was able to account -in separate steps- for the network structure and the oscillator intrinsic heterogeneity.
The effect of the network structure on the evolution of the individual oscillator phases was first accounted for
using the spectral properties of the network (under the assumption that the network graph Laplacian possesses a spectral gap).
In a second step, the effect of the heterogeneity in oscillator frequencies was accounted for by observing
(and then capturing) a strong correlation between (``excess") phase angles and
intrinsic oscillator frequency distribution.
Both steps were incorporated in the construction of a {\em lifting} and a {\em restriction} operator
(from coarse variables to detailed, fine scale state consistent realizations and vice versa).
These operators can then be linked, in the equation-free framework, with algorithms such as coarse fixed point
and coarse limit cycle computations, as well as with coarse projective integration, all of which were demonstrated.

Even though the individual oscillator dynamics and the network topology used for illustration here
were relatively simple, we are confident that the procedures demonstrated here can be extended to different,
more complex individual oscillator dynamics and for other types of networks, as long as they possess spectral gaps as well.
Extensions to networks of spiking neurons, which can also be considered as coupled oscillators -but with much more complex, and especially
{\em directional} coupling topologies- is probably
too ambitious with only these tools.
As pointed out in Ref.~\cite{Binz09topology}, ``{\em The information required to construct a detailed
and specific configuration of neocortex containing some $\mathit{10^{12}}$ connections exceeds by far
the roughly $\mathit{10^{8}}$ bits of information available in the genome for specification of the entire organism.
On these grounds alone it appears that nature's strategy for construction of the neocortex must depend
on the dynamic assembly of rather specific but simple modules}".
This reasoning supporting ``module simplicity and specificity"
provides hope and motivation for the deployment of reductionist
approaches in such systems \cite{Lain09stochastic}.

One particular direction of extension for which we are more confident,
and which we are currently pursuing,  is the study of diverse heterogeneity distributions.
In our illustrative example, the (intrinsic frequency) heterogeneity
distribution was simply a normal one (with different variances).
A single scalar quantity (the slope, $c$) of the correlation between
heterogeneity and system state was sufficient to improve our coarse
description here.
This slope is but the first nontrivial coefficient of an expansion
of the system state (here, the excess phases) as a function of a
random variable (here, the intrinsic frequency).
In effect, this is a ``one-term" polynomial chaos \cite{Ghan91stochastic} expansion of a function of a random variable
(the oscillator frequency) with a particular probability distribution.
It is straightforward to use different
expansions (depending on the distribution of the random variable,
different hierarchies of polynomials are applicable, see for
example the Askey scheme  \cite{Xiu02wiener--askey}); it is also
straightforward to use more than one term in the expansion in
a particular polynomial set if the correlation exhibits more
structure than the straight line we observed here.
This research avenue provides a direct link between existing and developing tools in the
study of uncertainty quantification (polynomial chaos approaches, and the associated collocation schemes)
with the study of coupled heterogeneous oscillator problems, even when heterogeneity arises in more than one properties of the
coupled system.
One particularly interesting direction for network dynamics arises
when the oscillator behavior depends crucially on the {\em degree}
of this oscillator in the overall network.
If correlations between node degrees and oscillator states quickly develop in system startup transients, the tools we outlined above may well serve
in successful coarse-graining of the overall network dynamics.

\appendix*
\section{\label{sec:ef}Outline of the Equation-Free Framework}
The Equation-Free (EF) approach to modeling and computation for complex/multiscale systems
has been developed for problems that can, in principle, be described at multiple levels.
In particular, it is applicable to systems for which
the evolution equations are available at a ``fine" (atomistic, microscopic,
individual-based) scale, while the equations for the ``coarse" (macroscopic, system level)
behavior, which is of interest, are not available in closed form.
We will illustrate the EF approach through a brief description of {\em coarse projective integration}
and coarse fixed point computation.
The system of interest is completely specified at any moment in time by a set of fine or
microscopic variables $U$.
We start with the assumption that an appropriate set of coarse variables $u$
(observables in terms of which closed equations can in principle be written
at the macroscopic level) have been selected.
We also assume that good {\em lifting} ($\mathcal{L}[\cdot$]) and {\em restriction}
($\mathcal{R}[\cdot]$) operators are available: the lifting creates fine scale initial
conditions consistent with prescribed values of macroscopic observables, while the restriction
obtains the values of the observables from a fine scale state.
These operators effectively ``translate" fine scale states to the corresponding coarse ones,
and coarse ones to consistent fine ones respectively.

For our illustrative example, the ``fine scale" state at a time instance $t_i = i \Delta t$ is the vector of phase angles of
all oscillators $U(t_i)$, while the corresponding set of coarse variable values is $u(t_i)$.
A coarse projective integration step consists of the following sub-steps:
\begin{enumerate}
\item
Lifting step: Start with initial condition $u(0)$ for the coarse macroscopic variables and
lift them to a consistent microscopic description:
$U(0)= \mathcal{L}[u(0)]$.
\item
Evolve the oscillators using $U(0)$ as the initial condition for their phases in the microscopic simulator
for a time $t_h$, long enough for the fast components of the dynamics to equilibrate, but short
compared to the slow (coarse) system time scales (see \cite{Kevr03equation-free}).
The final state of this step is $U(t_h)$.
\item
Evolve the microscopic variables, $U(0)$, for additional $k$ time steps, generating the values
$U(t_{i})$, $i=h+1$ to $h+k$, i.e., $U(t_{h+k})= \mathcal{E}_{t_{h+k}}[U(0)]$.
\item
Restriction step: Obtain the restrictions, $u(t_{i})= \mathcal{R}[U(t_i )]$, $i=h+1$ to $h+k$.
\item
Projective step: Estimate time derivatives from these restrictions,
$u(t_{i})$, $i=h+1$ to $h+k$, and use any numerical scheme (the simplest one would be forward Euler)
to ``project" the macroscopic variables ``into the future" over a time interval $p \Delta t$ to obtain $u(t_{h+k+p})$.
\end{enumerate}
One then uses these projected values of the coarse variables as the initial condition in repeating the overall procedure.

Using the lifting and restriction operations, and the fine-scale system simulator,
one can define a {\em coarse time-stepper} $\Phi_{\tau}$ (Eq.~\ref{eq:coarsets}) that takes as input
the coarse variables at a given time, $u(t)$, and outputs the coarse variables at a later time, $u(t + \tau)$.
\begin{equation}
u(t+\tau) = \Phi_{\tau} [u(t)] = \mathcal{R} [\mathcal{E}_{\tau} [\mathcal{L}  [u(t)]]].
\label{eq:coarsets}
\end{equation}

One can also use such a coarse time-stepper to find the coarse fixed points, $\widehat{u}$,
by solving Eq.~\ref{eq:fixed} for (in principle) any time $\tau$, using matrix-free implementations of
algorithms like Newton-Krylov-GMRES
\cite{Saad86gmres:,Kell95iterative} to iteratively solve sets of nonlinear equations.
\begin{equation}
\widehat{u} = \Phi_{\tau} [\widehat{u}].
\label{eq:fixed}
\end{equation}

With the help of coarse Poincar\'e maps, one can solve a similar equation to find a (coarse) point on the
(coarse) limit cycle, $\widetilde{u}$, as well as its period, $T$.
\begin{equation}
\widetilde{u} = \Phi_{T} [\widetilde{u}].
\label{eq:limit}
\end{equation}
Matrix-free implementations of eigensolvers (e.g. matrix-free Arnoldi procedures \cite{Leho98arpack}) can (and have been) used
to characterize the coarse linearized stability of coarse fixed points and limit cycles \cite{Sama08newton-krylov,Siet03enabling,Lust98adaptive}.
\begin{acknowledgments}
This work was partially supported by the US DOE (DE-SC0002097) and by DTRA (HDTRA1-07-1-0005).
\end{acknowledgments}


%

\end{document}